\documentclass[conference]{IEEEtran}
\ifCLASSINFOpdf
\else
\fi
\usepackage[cmex10]{amsmath}
\usepackage{amssymb}
\usepackage{mathtools}
\newcommand{\inlineeqnum}{\refstepcounter{equation}~~\mbox{(\theequation)}}
\usepackage{microtype}
\usepackage{paralist}
\usepackage{multirow}
\usepackage{booktabs}
\usepackage{xcolor}
\usepackage{flushend}

\begin{document}
%
\title{Melody-Conditioned Lyrics Generation with SeqGANs}

\author{\IEEEauthorblockN{Yihao Chen}
\IEEEauthorblockA{Center for Music Technology\\
Georgia Institute of Technology\\
Email: ychen3237@gatech.edu}
\and
\IEEEauthorblockN{Alexander Lerch}
\IEEEauthorblockA{Center for Music Technology\\
Georgia Institute of Technology\\
Email: alexander.lerch@gatech.edu}
}


\IEEEpeerreviewmaketitle

\maketitle

\begin{abstract}
Automatic lyrics generation has received attention from both music and AI communities for years. Early rule-based approaches have~---due to increases in computational power and evolution in data-driven models---~mostly been replaced with deep-learning-based systems. Many existing approaches, however, either rely heavily on prior knowledge in music and lyrics writing or oversimplify the task by largely discarding melodic information and its relationship with the text. We propose an end-to-end melody-conditioned lyrics generation system based on Sequence Generative Adversarial Networks (SeqGAN), which generates a line of lyrics given the corresponding melody as the input. Furthermore, we investigate the performance of the generator with an additional input condition: the theme or overarching topic of the lyrics to be generated. We show that the input conditions have no negative impact on the evaluation metrics while enabling the network to produce more meaningful results.
\end{abstract}
\begin{IEEEkeywords}
lyrics generation, melody conditioned, deep learning, SeqGAN, music information retrieval.
\end{IEEEkeywords}
\section{Introduction}
Lyrics writing is a crucial part of the process of songwriting, and good lyrics contribute to expressiveness and influence the emotional valence of the music \cite{ali_songs_2006}. Writing lyrics from scratch, however, does not come easily to everybody. The task is comparable in its complexity to poetry writing with similar demands on expressiveness, conciseness, and form. Additional constraints due to melodic properties require basic music understanding and complicate the task even more. Thus, automatic lyrics generation is a useful and important task which aims at providing musicians with inspirations for song writing. Due to the similarities of the tasks, many approaches for lyrics generation are inspired by text generation systems. Early approaches used to be rule-based text generation methods with limited musical information included \cite{gonccalo2007tra,  DBLP:journals/jagi/Oliveira15/SemanticGeneration}. However, those systems relied heavily on prior knowledge in lyrics writing and music to define fitting rules. Driven by advancements in text generation through the surge of neural networks and deep learning, more data-driven lyrics generation models were proposed \cite{watanabe-etal-2018-melody, potash-etal-2015-ghostwriter}. However, similar to rule-based systems, existing learning-based methods only incorporate very limited melodic information \cite{Fan2019Seq2SeqChinese, Nikola2020RapGen} and are therefore not able to generate lyrics that fit given melodies. The few systems incorporating melodic information require complex pre-processing and representation of both lyrics and melody data \cite{watanabe-etal-2018-melody}. 

To address these limitations, we propose a system that implicitly learns the interrelation of lyrics and melody and uses such knowledge to generate lyrics given a specific input melody. The end-to-end system does not require prior knowledge in lyrics writing and music composition or complex data pre-processing. Inspired by the success of Generative Adversarial Networks and their variants \cite{goodfellow2014generative, DBLP:conf/aaai/YuZWY17/seqGAN}, we propose a deep-learning approach to lyrics generation using SeqGAN conditioned on melody input. The main contributions of this paper are:

\begin{enumerate}[(i)]
    \item   the presentation of an end-to-end system with adversarial training for lyrics generation conditioned on the musical melody,
    \item   an ablation study on the impact of the melody conditioning on the generated lyrics, and
    \item   a preliminary study of the system performance with an additional theme constraint.
\end{enumerate}

The remainder of this paper is structured as follows: Section~\ref{sec:lit} provides an overview of lyrics generation related researches. Section~\ref{sec:method} introduces the formulation of melody conditioned lyrics generation problem and proposed lyrics generation models. Section~\ref{sec:results} describes 4 experiments conducted for evaluating the models and discusses the results. Lastly, Sect.~\ref{sec:conclusion} concludes the paper.

A web-based demo of the presented system can be accessed online.\footnote{https://lyrics-lab.herokuapp.com/}

\section{Related work}\label{sec:lit}
Generative models have been an active area of research for many years. Text, as one of the most ubiquitous content domains, has become an increasingly popular target for researchers in this area. Historically, statistical language models had been heavily used, with the n-gram language model as one of the most widely adopted generative systems \cite{Jurafsky:2009:SLP:1214993}. They were, for example, used for dialogue generation and machine translation \cite{oh2000stochastic, Liu1198861ngramtranslation}. Despite their general usefulness, n-gram models are criticized for being sensitive to cross-domain corpora and having false independence assumptions \cite{Rosenfeld2000SLM}. For instance, the assumption that the next word in a sentence depends only on the identity of the n-1 preceding words is inaccurate for natural languages \cite{Chomsky1956ThreeModels, chomsky2002syntactic}.

In the past decade, neural networks have been increasingly applied to the automatic text generation task. Recurrent Neural Networks (RNNs), in particular, have been identified as powerful architectures for modeling language and capturing sequence dependency \cite{mikolov2010recurrent,  Hocheriter1997LSTM}.  
Text generation models based on recurrent architectures can be roughly grouped into three categories, supervised learning, unsupervised learning, and Reinforcement Learning (RL). In supervised learning, the Maximum Likelihood Estimation (MLE) is a widely-used training paradigm since it approaches the generation problem as a sequential multi-label classification problem and directly optimizes the multi-label cross-entropy \cite{Williams1989RNN, LuDBLP:journals/corr/abs-1803-07133NLG}. Each label in the context of text generation is a word or syllable in the vocabulary of the corpus. MLE tends to be more robust and converges faster than other algorithms during training \cite{Karpathy2017, pmlr-v70-hu17e}. However, it suffers from oversimplified training objectives, exposure bias, and loss mismatch \cite{DBLP:conf/emnlp/LiMSJRJ17, huszar2015not}. As the objective of MLE training is merely estimating conditional probabilities of sequences, the resulting texts tend to be unnatural and not interesting \cite{Liu1198861ngramtranslation}. Moreover, the cross-entropy loss used for MLE cannot guarantee a satisfying text output  as it is not specifically designed for text generation \cite{DBLP:conf/emnlp/LiMSJRJ17}.

Different from the MLE-based methods, models such as the Variational Autoencoder (VAE) take an unsupervised or semi-supervised approach to content generation. The VAE can extract a latent and continuous representation of text that can be sampled and interpolated for generation, allowing for variety in the results \cite{bowman-etal-2016-generating}. Nevertheless, the VAE seems to be unable to accurately model complex, large scale datasets unless other approaches such as adversarial training or supervised features are employed \cite{Dosovitskiy2016GenImage, lamb2016discriminative}. More recently, researchers have started to investigate the applicability of Generative Adversarial Networks (GANs) in text generation \cite{goodfellow2014generative}, although the direct application is not easy since GANs by premise only support differentiable operations, which excludes text sampling. Later, the Sequence Generative Adversarial Network (SeqGAN) adopted the policy gradient from RL to bypass the problem of non-differentiability \cite{DBLP:conf/aaai/YuZWY17/seqGAN} and enabled the GAN architecture to handle text generation. 

The variety of text generation tasks ranges from poetry to speech draft generation \cite{zhang2014chinese, DBLP:conf/aaai/YuZWY17/seqGAN}. Lyrics generation inherits complexities in syntax and semantics from text generation. In addition to these complexities, other constraints such as syllabic structures, rhyme, pitch value, and rhythm should also be taken into consideration. To generate lyrics that match these constraints and requirements and exhibit meaning, efforts have been made through both rule-based and learning-based approaches. Linguistic analysis and predefined generation strategies have been adopted for automatic lyrics generation \cite{DBLP:journals/jagi/Oliveira15/SemanticGeneration, RamakrishnanA_AutoTamilGen(Rule)}. However, the quality of generated lyrics from those systems heavily relies on knowledge in linguistics and lyrics writing for exploitative analysis and good rule design.

Therefore, learning-based approaches which require little or no human prior knowledge drew researchers' attention, and several aspects of lyrics generation have been studied. Zhang et al.\ analyzed syntactic formation in terms of word ordering in text generation and proposed a learning-guided word search framework for imposing word ordering on the generated outputs \cite{doi:10.1162/Syntaxtextgen}. Lu et al.\ have shown the ability of a Seq2Seq RNN encoder to learn syllabic structure for Chinese lyrics generation \cite{10.1007/978-3-030-29894-4_20:SyllableGen}. Potash et al.\ illustrated that an LSTM is effective in generating various rap lyrics while preserving the rapper's style \cite{potash-etal-2015-ghostwriter}. Despite all these approaches, it is noteworthy that only few incorporate any melodic information and are thus unable to generate melody-matching lyrics.

\section{Proposed Method}\label{sec:method}
This section introduces the melody-conditioned lyrics generation problem and presents the generative systems investigated in this study: two baseline models and three generative SeqGAN architectures. 

\subsection{Problem Formulation}
The melody-conditioned lyrics generation problem can formally be described as follows. The corpus $\mathcal{C} = (\mathcal{L}, \mathcal{M})$ consists of paired lyrics sequences  $\mathcal{L} = \{L_1, L_2, \hdots, L_n\}$ and melody sequences $\mathcal{M} = \{M_1, M_2, ..., M_{n}\}$. The lyrics $L_i = \{l_1, \hdots, l_{T}\}$ and the melody sequences  where $M_i = \{m_1, \hdots, m_{T}\}$ correspond to each other so that each word or syllable $l_i$ in $L_i \in \mathcal{L}$ has a corresponding token $m_i$ in $M_i \in \mathcal{M}$ and $\forall l_i \in \mathbf{L_\mathrm{vocab}}$, $\forall m_i \in \mathbf{M_\mathrm{vocab}}$. The information contained in $m_i$ might vary depend on the representation for the melody. In this paper, $m_i$ represent a tuple of pitch value and note duration.

The generative model $G_{\theta}$ with parameters $\theta$ is trained on the corpus to generate a sequence $L'_i = \{l'_1, \hdots,l'_{T}\}$ given the conditioning information from $M'_i = \{m'_1, \hdots,m'_{T}\}$, where $\forall l'_i \in \mathbf{L_\mathrm{vocab}}$ but not necessary $\forall m'_i \in \mathbf{M_\mathrm{vocab}}$ since the input notes during the inference phase might not have been seen during the training.

\subsection{Baseline Systems}
The N-gram language model is a standard baseline model for text generation tasks \cite{Jurafsky:2009:SLP:1214993}; here, the probability $P(l_1, ..., l_n)$ of generating a text token sequence $(l_1, l_2, ..., l_n)$ is computed by using the preceding tokens' conditional probabilities $P(l_n| l_{1}^{n-1})$ approximated with 
 \begin{equation}\label{eq:n-gram}
     P(l_n|l_1^{n-1})\approx P(l_n|l_{n-N+1}^{n-1})
 \end{equation}

This model is called an n-gram model, and sequences are generated by sampling sequential tokens maximizing the probability above. A bi-gram model is Eq.~(\ref{eq:n-gram}) with $N = 2$

We compute our results for two baseline models: an unconditioned bi-gram and a melody-conditioned bi-gram model (MC bi-gram). The MC bi-gram additionally incorporates the current note information in the melody to model the following probability: $P(l_n|l_1^{n-1}, m_1^{n})\approx P(l_n|l_{n-1}, m_{n})\inlineeqnum\label{eq:mc bi-gram}$.

In case a sampled $(l,m)$ tuple does not occur in the training corpus, the next token is sampled using unconditioned bi-gram probability.

\subsection{SeqGAN Generators as LSTM Models}
There are two training phases in SeqGAN models: MLE and adversarial training phase. During MLE training phase, the generator is trained to predict the following token given a preceding sequence. Given that LSTM is the backbone of the generator in SeqGAN architecture, training generators with MLE paradigm equals training LSTM models. The performance of the LSTM models on the lyrics generation task can be evaluated with the performance of the generator after MLE training. In the following experiments, we will be using SeqGAN (MLE), MC-SeqGAN (MLE), and TMC-SeqGAN (MLE) to represent unconditioned LSTM, MC-LSTM, and TMC-LSTM. Details about these conditioned SeqGANs will be discussed in the following sections.

\begin{figure}[t]
 \centerline{
 \includegraphics[scale= 0.8]{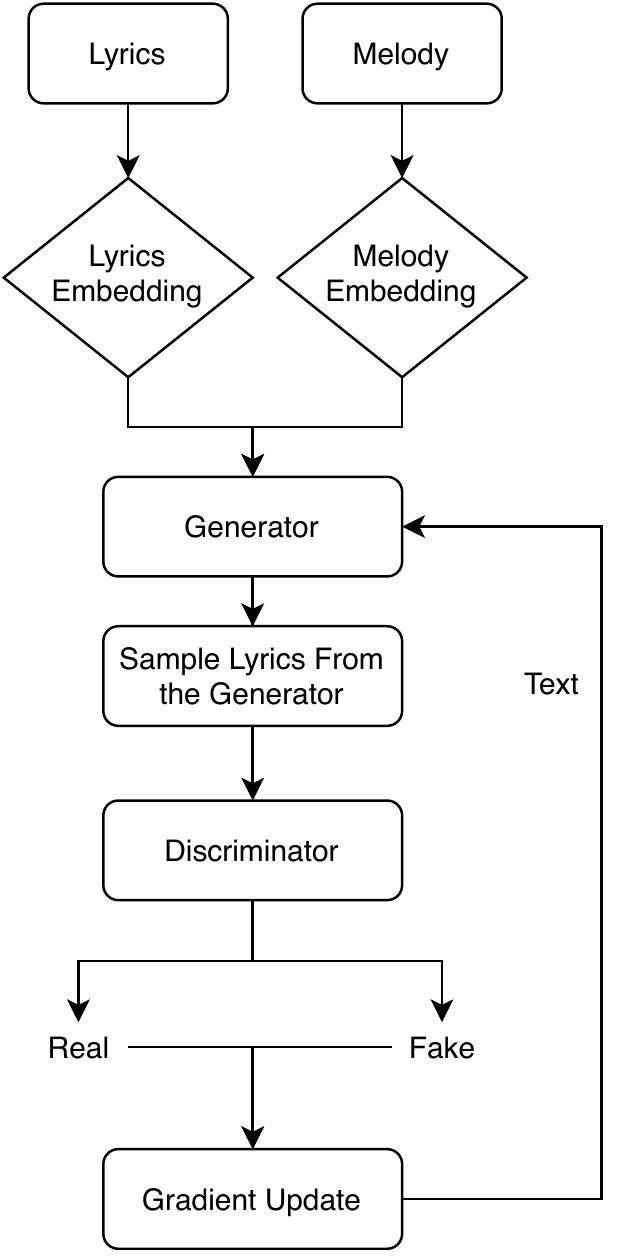}}
 \caption{Flow chart of the Melody-Conditioned SeqGAN}
 \label{fig:example}
\end{figure}

\subsection{Sequential Generative Adversarial Network}
SeqGAN utilizes the policy gradient from RL to maximize the expected reward for generating a sequence:
\begin{eqnarray}\label{PG Reward}
    J(\theta) &=& \mathbf{E} [R_{T}|s_0, \theta]\nonumber\\ 
    &=& \sum_{m_1 \in \mathbf{M_{vocab}}} G_{\theta}(m_1|s_0) \cdot Q_{D_{\phi}}^{G_{\theta}}(s_0, m_1),
\end{eqnarray}
where $R_{T}$ is the reward for a complete sequence, and $Q_{D_{\phi}}^{G_{\theta}}(s_0, m_1)$ is the action-value function of generating a sequence $m_1$ given start state $s_0$, generator $G_{\theta}$, and discriminator $D_{\phi}$ \cite{DBLP:conf/aaai/YuZWY17/seqGAN}.

One problem with this strategy is that the reward value is only provided for a finished sequence from the discriminator. The intermediate results, however, are also important in our case; for instance, a good sub-sequence could end in a bad finish. Hence, a Monte Carlo search with a rollout policy $G_{\beta}$ is included for our SeqGAN to sample remaining tokens for an incomplete sentence to compute an immediate reward for a pseudo-finished sequence. This means that given a partially generated sequence with a length smaller than the maximum length, it will be expanded to the maximum length with Monte Carlo sampling for the policy gradient computation. After the gradient update for the generator, the discriminator is re-trained with real data from the corpus and fake data from the generator.

\subsection{Melody-Conditioned SeqGAN}
To include musical information from the melodies, a melody embedding is added to the generator to combine it with the lyrics representation. The trainable lyrics and melody embeddings are concatenated and fed into the generator. During sampling at timestep $t$, the note $m_{t+1}$ is referenced to generate the lyrics $l_{t+1}$; during inference, a melody line is taken as an input and a lyrics line is generated. Figure~\ref{fig:example} visualizes the training process of the resulting MC-SeqGAN.

\subsection{Themed MC-SeqGAN}
\begin{figure}
 \centerline{
 \includegraphics[scale = 0.8]{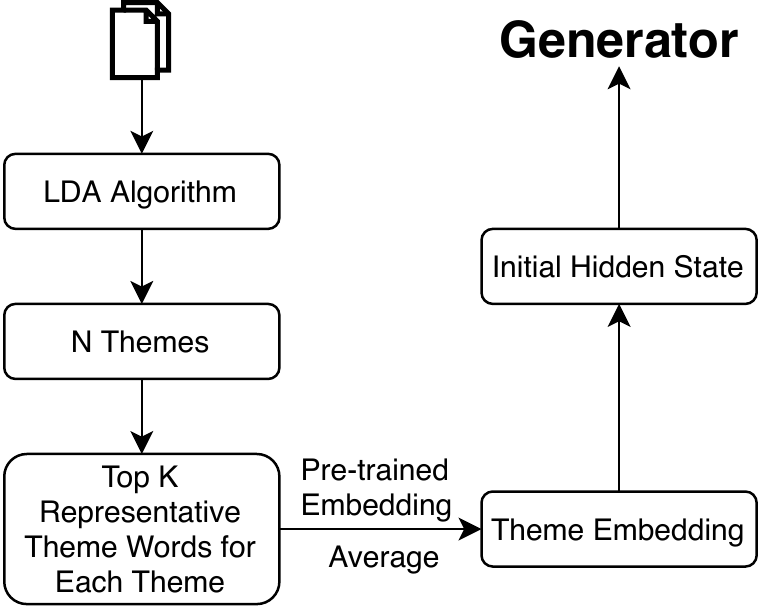}}
 \caption{Theme extraction and representation in TMC-SeqGAN training.}
 \label{fig:themed}
\end{figure}

Lyrics often can be grouped into themes such as love, friendship, etc. Learning theme information enables the model to generate lyrics within a specified theme or topic domain. Our proposed work flow of specifying a theme is shown in Fig.~\ref{fig:themed}. Inspired by neural architectures for image captioning, where the image representation is compressed and passed to the RNN layer in the following sequence generation model as the initial hidden state \cite{You_2016_CVPRImageCaptioning}, we extract a representation of themes from the lyrics corpus and feed the theme information of each line into the generator of the MC-SeqGAN. The theme extraction is described in Sect.~\ref{sec:data}. 

\section{Experiments and Results}\label{sec:results}
We provide a quantitative evaluation of three aspects our lyrics generators,
\begin{inparaenum}[(i)]
    \item   the quality of the generated lyrics (n-gram, MC-n-gram, SeqGAN, MC-SeqGAN, TMC-SeqGAN),
    \item   the alignment between the generated lyrics and input melody (MC-SeqGAN, TMC-SeqGAN), and
    \item   the theme match of the generated lyrics in the case of the TMC-SeqGAN. 
\end{inparaenum}
Although we use established metrics, the power of the used objective metrics to predict perceptual quality is unproven.
Therefore, we additionally provide examples for generated lyrics and discuss them in a qualitative evaluation section.


\subsection{Data and Input Representation}\label{sec:data}

The dataset used for this task is the Lyrics-Melody Dataset created by Yu et al\cite{Yu2020LCMelody}., which contains aligned lyrics-melody pairs extracted and processed from the Lakh\_full MIDI Dataset \cite{raffel2016learning}. The dataset contains lyrics and melodies for 7,998 songs with the lyrics pre-tokenized at the syllable level. There are 20,934 unique syllables and 20,268 unique words in the dataset. The midi pitch of the melody notes ranges from 21 to 108, and the allowed note duration ranges from a 16th note to 8 times the length of a whole note. For all the following experiments, the input lyrics and melodies are organized at line level, meaning that each input entry is one line of lyrics and accompanying melody. 

The lyrics are encoded into discrete indices using a syllable-to-index mapping extracted from the corpus with every syllable having a unique index. The \{pitch, duration\} tuples are used for melody representation and they are encoded with note-to-index mapping. Each sequence begins and ends with \texttt{<START>} and \texttt{<END>} tokens, respectively, and all sequences are padded to the maximum sequence length observed from the dataset with a \texttt{<PAD>} token to enable batched training with the RNN. Only tokens between \texttt{<START>} and \texttt{<END>} are used for evaluation purposes.

\begin{figure}
 \centerline{
 \includegraphics[trim=0cm 0.5cm 0cm 0.7cm, width = \columnwidth]{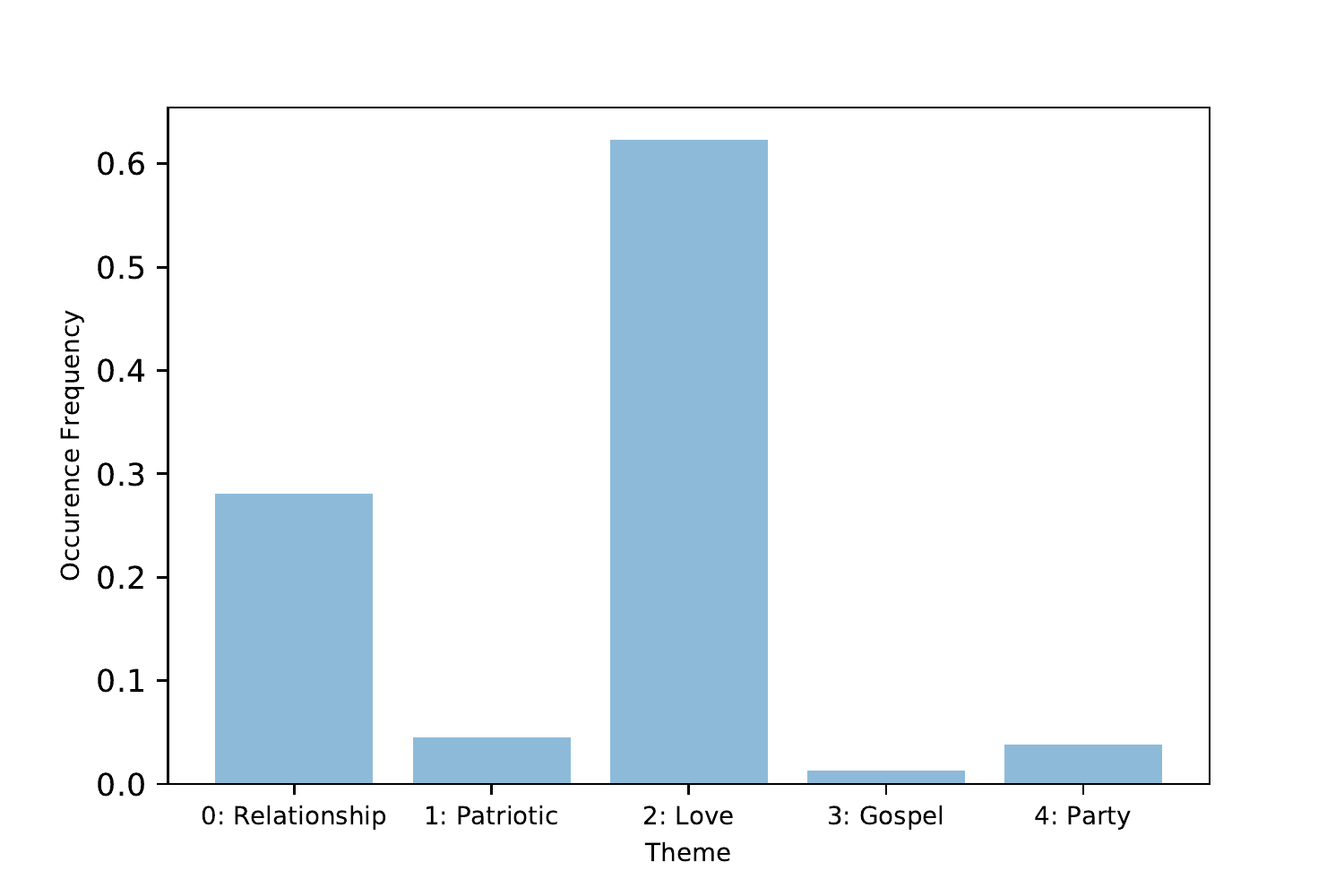}}
 \caption{Distribution of theme classes extracted from the corpus using LDA. Themes~0 is about relationship; theme~1 is about patriotism; theme~2 is about love; theme~3 is about gospel; theme~4 is about party.}
 \label{fig:distribution}
\end{figure}

As the data does not come with theme annotations, there is no absolute ground truth to train and evaluate the TMC-SeqGAN model. To the best of our knowledge, there is no public dataset containing lyrics, melodies, and themes. Therefore, we generate a pseudo ground truth with the Latent Dirichlet Allocation (LDA) algorithm, a widely used probabilistic topic modeling method \cite{blei2003LDA}.
LDA classifies the corpus into a specified number of themes and generates a set of representative words for each theme identified. Every line in one song is assigned the same theme. Stop-words are removed from the corpus before performing LDA. The number of themes $\mathcal{N}$ to be classified is a hyperparameter of LDA, and different values for $\mathcal{N}$ need to be explored by comparing the coherence score and the perplexity of the theme model \cite{Newman2010TopicCoherence}. Furthermore, the distribution of the resulting themes should not be extremely skewed. After comparison, we found $\mathcal{N}=5$ to be a fitting number of themes for our lyrics corpus. However, as shown in Fig.~\ref{fig:distribution}, a skewness towards Theme~2 is still present in the resulting themes. Furthermore, the LDA results show a small overlap between Themes~0 and 2 as well as between Themes~3 and 4.
One limitation of the LDA algorithm is that the name for each theme must be manually assigned. We tried to verbalize each theme with one or two words. We categorized Themes~0 as ``relationship'' since the cluster's top representative words contain ``baby'', ``away'' and ``go'', while Theme~2 is related to ``love'' with representative words including ``love,'' ``never,'' ``together,'' and ``heart.'' Theme~4 is categorized as ``party'' with representative words like ``party,'' ``dance,'', and ``music,''. Themes~1 and 3, however, can not be easily verbalized since they contain a mixture of words not easily categorized into one theme. Despite these challenges we assign Theme~1 the label ``patriotic'' according to words ``America'' and ``country'' and Theme~3 ``gospel'' due to the words ``heaven'' and ``Jesus'' for reference purposes. After theme extraction, the average of the word vectors of top $\mathcal{K}=10$ representative words for each theme is regarded as the theme embedding and is calculated using the \textit{Fasttext} 1 million English word vectors, trained on \textit{Wikipedia 2017}, the \textit{UMBC} webbase corpus, and the \textit{statmt.org} news dataset\cite{mikolov2018advances}\footnote{https://fasttext.cc/docs/en/english-vectors.html Last access: 2020/10/25}. This theme embedding is used to initialize the hidden state of the generator.

For all experiments, the processed and encoded corpus was divided into training set (80\%), validation set (10\%) and test set (10\%). Lines of lyrics and melodies are grouped in pairs and then shuffled. For Experiment~3, the theme embedding is added to lyrics-melody group before shuffling.

\subsection{Evaluation Metrics}

The quality of the generated lyrics is evaluated with the BLEU score. The BLEU score is a standard quantitative metric often used for machine translation and text generation \cite{Papineni:2002:BMA:1073083.1073135/BLEU}. We used the bi-gram cumulative BLEU score (BLEU2) and 4-gram cumulative BLEU score (BLEU4) in this study. These scores are calculated for each line of generated lyrics with regard to the lyrics in the validation set. Both the average and the standard deviation of the calculated scores are reported.

The alignment of the generated lyrics with the input melody is tested with the alignment ratio, the number of lyrics lines where the number of syllables equals the number of notes over the total number of lyrics lines. 

Investigating the discriminator performance gives us additional insights into our model. The discriminator is evaluated with the metrics precision, recall, and F1 score. Note that in the case of the discriminator, the reported scores are based on the validation set.

Last but not least, the themes are evaluated with the accuracy. To ensure that the metric is not biased due the skewness of the data, the macro average F1 (the average over all classes' F1 scores) and micro average F1 (the global F1 score over all predictions) are reported.

\subsection{Experimental Setup}

\begin{table}
  \caption{Generators' Performance Comparison. MLE and Adv represent two training phase in SeqGAN training, and the MLE training is followed by adversarial training. The average BLEU scores of generated lyrics are followed by their standard deviations. The higher the average BLEU score, the better lyrics' general quality. The lower the standard deviation, the better the consistency among generated lyrics' quality}
  \label{tab:G}
  {%
    \begin{tabular*}{\columnwidth}{l  @{\extracolsep{\fill}}cc}
    \toprule
    \textbf{Model} & \textbf{BLEU2} & \textbf{BLEU4}\\
    \midrule
    Bi-gram & 0.522 $\pm 0.173$ & 0.239 $\pm 0.106$ \\
    MC Bi-gram & 0.537 $\pm 0.156$ & 0.177 $\pm 0.122$ \\
    \midrule
    SeqGAN (MLE)& 0.830 $\pm 0.168$ & 0.445 $\pm 0.187$  \\
    SeqGAN (Adv) & {0.904 $\pm 0.095$} & 0.421 $\pm 0.147$  \\
    \midrule
    MC-SeqGAN (MLE) & 0.830 $\pm 0.168$ & 0.431 $\pm 0.177$  \\
    {MC-SeqGAN (Adv)} & 0.868 $\pm 0.143$ & 0.439 $\pm 0.172$  \\
    \midrule
    TMC-SeqGAN (MLE) & 0.829 $\pm 0.165$ & 0.446 $\pm 0.185$  \\
    {TMC-SeqGAN (Adv)} & 0.873 $\pm 0.140$ & {0.487 $\pm 0.196$}  \\
  \bottomrule
\end{tabular*}%
}
\end{table}

\subsubsection{Experiment 1} 
The generated lyrics' quality is compared among the bi-gram baselines, SeqGAN, MC-SeqGAN, and TMC-SeqGAN. For the baseline models, the transition probabilities are directly extracted from lyrics and syllable-note tuples, respectively, to create unconditioned and conditioned bi-gram models. 

The generator in the SeqGAN model contains one embedding layer of size 128 for input representation learning, a single one directional LSTM layer of hidden size 128 for learning sequential parameters, and a fully-connected layer followed by a log softmax function for mapping the output to the vocabulary space. The difference between SeqGAN and MC-SeqGAN is only the dimension of the embedding layer, where the conditioned model concatenates the lyrics and melody embeddings while the unconditioned architecture one only has lyrics embedding. There are around 1.1M and 1.2M trainable parameters for unconditioned and conditioned SeqGANs.

The discriminator has the same architecture in all models. It has one embedding layer followed by sequential CNN layers based on evidence that CNNs are efficient in token sequence classification \cite{kim-2014-convolutional, DBLP:journals/corr/ZhangL15/CNNText}. Following the CNN layers are two fully connected layers with one dropout layer for generating classification decisions. There are around 0.6M trainable parameters in discriminators.

For all models based on the SeqGAN, an MLE pretrain phase is applied before adversarial training to help the stability and convergence of the generator and discriminator. For the generator, the pre-training is a supervised training with tokens $\{t_1, t_2, \hdots, t_{n-1}\}$ as the input and $\{t_2, t_3, \hdots, t_{n}\}$ as the target, while for the discriminator the procedure is the same as in the adversarial phase, where real data and fake data are provided for classification. All generators are pre-trained for 120 epochs using Adam optimization and the discriminator is pre-trained for 50 epochs using the Adagrad optimizer. As mentioned before, the pretrained generators can also be regarded as trained LSTM models. During the adversarial training phase, the generator and the discriminator are trained for 50 rounds. Within each round, the generator is trained with 1 epoch with 10 mini-batches, and then the discriminator is trained for 1 epoch with the whole training set to ensure fast convergence and stable training. The models with the best scores on the validation set are selected for subsequent analysis.

\begin{figure}
\vspace{-5mm}
 \centerline{
 \includegraphics[width = \columnwidth, trim = 0cm 2cm 0cm 1cm]{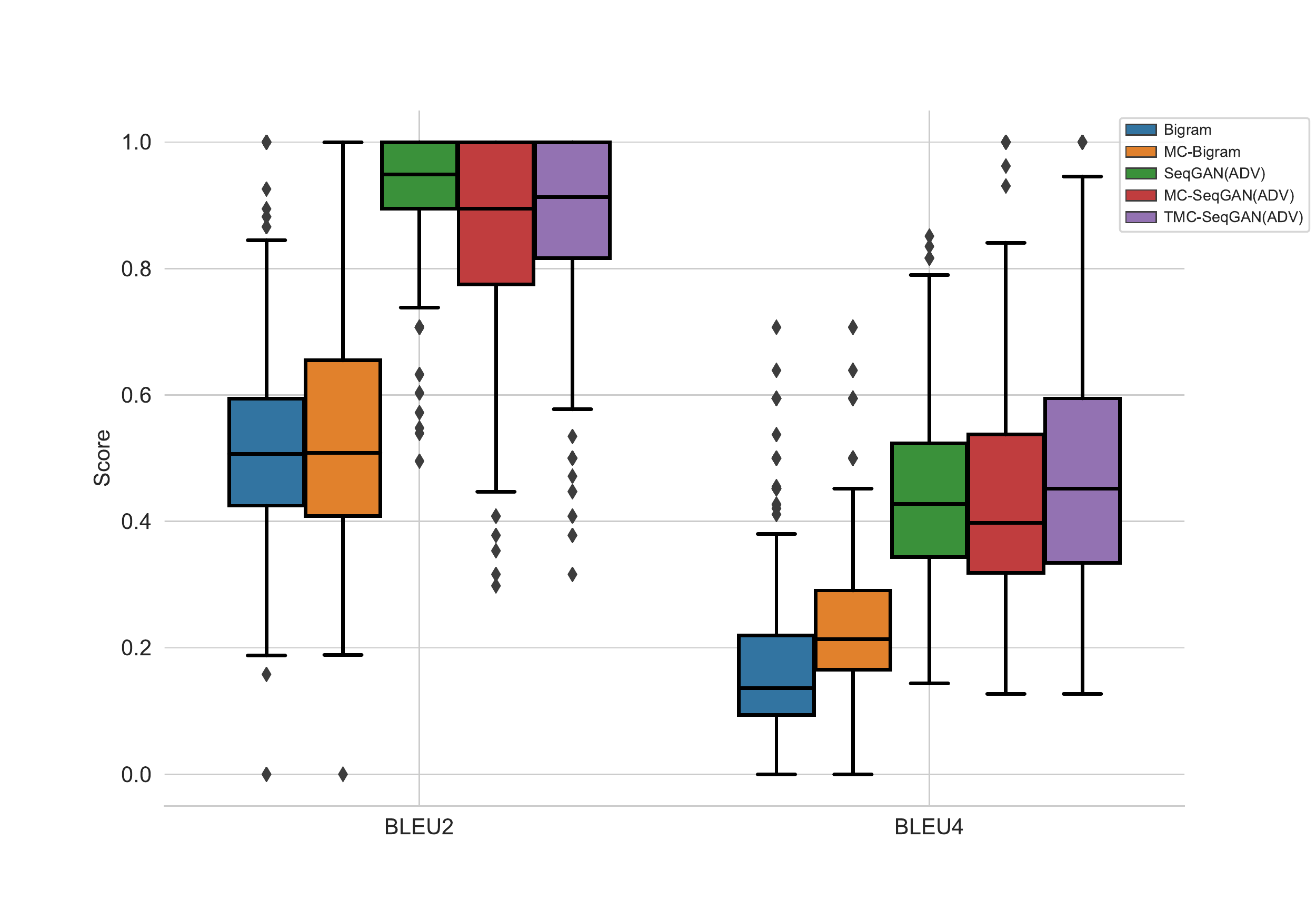}}
 \caption{The box plot for generators' performance. Different colors indicate different models. SeqGAN variants outperform the baseline models and additional constraining information do not significantly impact SeqGAN models' performance. }
 \label{fig:boxplot}
\end{figure}

\subsubsection{Experiment 2} 
The alignment between lyrics and the melody is evaluated for both MC-SeqGAN and TMC-SeqGAN, using the alignment ratio discussed above. The models will first perform inference to generate a corpus of lyrics given the melodies in the test set, and then we computed the alignment ratio regarding those melodies. MC bi-gram is not evaluated since it is enforced that the sampling ends when the melody ends, and SeqGAN is not evaluated as there is no paired melody to compare the alignment with. 

\subsubsection{Experiment 3} 
The themed MC-SeqGAN, an extension of the MC-SeqGAN, allows setting specific themes for the generated lyrics. We evaluate the success of this additional theme conditioning by comparing the themes of the generated lyrics~---extracted with the LDA model reference above---~with the expected themes using the macro and micro average F1.

\subsubsection{Experiment 4}
Since the above metrics do not sufficiently reflect the perceptual quality of the lyrics and the reflection of themes, a non-exhaustive qualitative evaluation is included. We randomly sampled 64 lines of lyrics from SeqGAN, MC-SeqGAN, and TMC-SeqGAN models and evaluate the quality of the lyrics, the relation to the input melody, and whether the lyrics reflect the specified the themes.


\subsection{Results and Discussion}

\subsubsection{Experiment 1} 
Table~\ref{tab:G} and Fig.~\ref{fig:boxplot} show that the performance of the baseline systems is not satisfactory, which is as expected as they only produce sequences according to extracted transition probabilities witnessed in the training corpus. An additional problem is the sensitivity to the corpus (compare \cite{Rosenfeld2000SLM}), i.e., the n-gram models cannot perform well if the test corpus contains too many unseen lyrics or lyrics-melody pairs, respectively. 
The quality of generated lyrics significantly increased with MLE training and is further improved after adversarial training, with the exception of a small decrease in BLEU4 of SeqGAN. The increases become more significant with more conditioning information. We also observe a trend of decreasing standard deviation of the BLEU scores. These results suggested that the proposed models are capable of modeling short lyrics-melody phrases, however, there is still space for improving the modeling of longer phrases since the BLEU4 scores are all much lower than BLEU2. Compared to the original SeqGAN, the performances of the conditioned SeqGAN variants are not significantly impacted with additional melody and theme constraints, suggesting the possibility of incorporating more customaizable constraints to lyrics generation without sacrificing quality.

\subsubsection{Experiment 2} 
The alignment ratio is only evaluated for the conditioned SeqGAN models as explained above. The alignment ratio of both the MC-SeqGAN and TMC-SeqGAN models equals 1, meaning that every line of generated lyrics is perfectly aligned with the input melody. The alignment between the lyrics and melody is successfully and implicitly learned by the model during the training.


\begin{figure}
 \centerline{
 \includegraphics[trim = 0.5cm 0.5cm 0cm 0.5cm, width = \columnwidth]{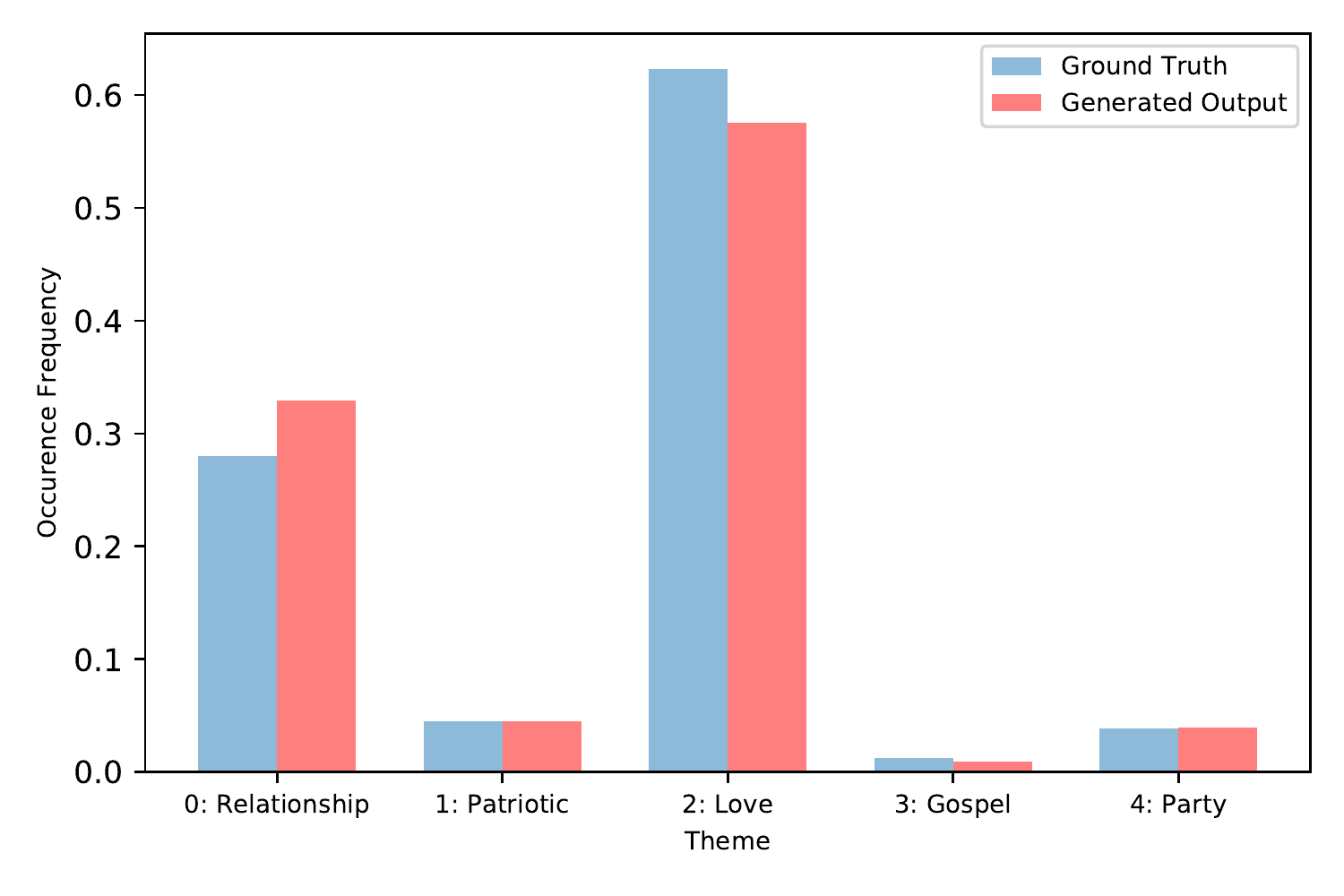}}
 \caption{Distribution of themes in extracted from the test set and generated lyrics.}
 \label{fig:prediction}
 
\end{figure}

\subsubsection{Experiment 3} 
The TMC-SeqGAN performs on a par with the MC-SeqGAN model in terms of quality and alignment ratio of the generated lyrics as discussed above. We have yet to investigate, however, whether the generated lyrics reflect the intended themes.
The macro average F1 for the themes as extracted with the LDA model is 0.60, and the micro average F1 is 0.74, which indicate that our model is not simply learning the skewness of the themes. The distribution of the themes in generated lyrics and ground truth is shown in Fig.~\ref{fig:prediction}. The results above show that the model can learn how to generate lines of lyrics given theme specification. 

\begin{figure}[t!]
 \centerline{
 \includegraphics[trim = 0.5cm 0.5cm 0cm 0.5cm, width = \columnwidth]{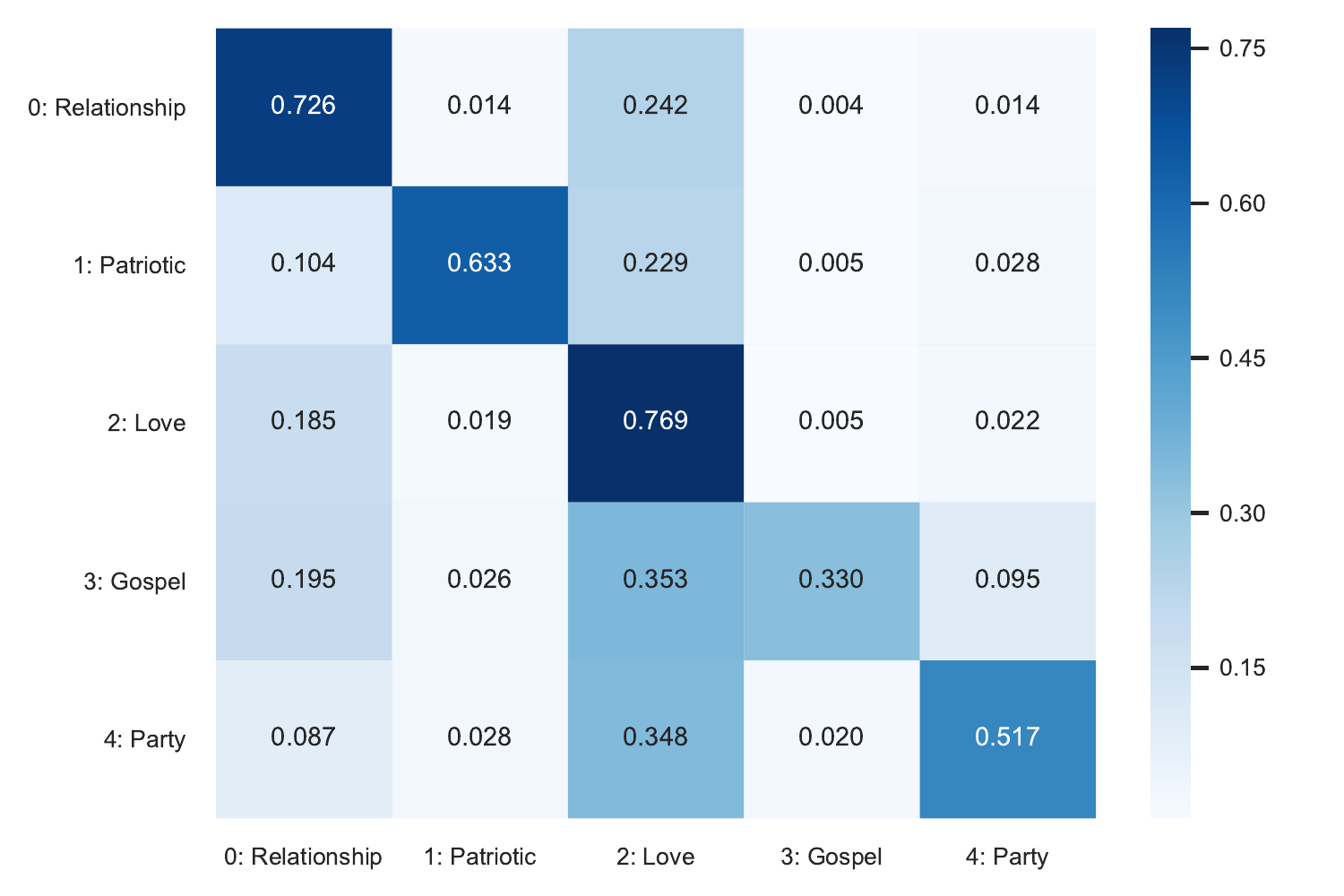}}
 \caption{The confusion matrix of theme predictions. The y-axis is the ground truth labels, while the x-axis is the predictions. The scale on the right is the occurrence frequency. The darker the color, the higher the frequency}
 \label{fig:Confusion_Matrix}
\end{figure}

However, as shown in Fig~\ref{fig:Confusion_Matrix}, while lyrics generated with Themes~0 and 2 generally exhibit focus on relationship and love, the results of Theme~1, 3, and 4 are ambiguous. These results match the observations made in section \ref{sec:data} where those themes were hard to verbalize, and they highlight the necessity of having a clear definition of themes. Besides, the uneveness among labels can also lead to the poor performance on those three themes. Moreover, there are problems with lyrics of Themes~0 and 2: some lyrics like ``You love to pretend'' are not necessarily a lyrics about love, but will still be categorized as a love song due to the presence of the word ``love.'' This is a limitation of our assumption about theme extraction that a set of representative words is sufficient for modeling a theme. 

It is important to remember that the extracted themes are only a pseudo ground-truth and LDA is not a perfect theme extraction algorithm. It models themes of the documents with keyword probabilities. Sometimes keywords do not illustrate the theme well due to the use of, e.g., metaphors. Besides, as is discussed in section \ref{sec:data}, verbalizing the themes after extraction can be harder than assigning lyrics into pre-named theme classes since the representative words can be too general or abstract to be categorized, like ``one'', and ``everybody''. Learning and evaluating themes annotated by humans would be better in this case. In future work, we want to explore the theme conditioning on an evenly distributed dataset, and ideally a manually annotated one. 

Additionally, it has been shown that GAN models suffer from mode collapse \cite{CheLJBL17ModeCollapse}. They will learn to generate few unique data that are been accepted by the discriminator, while the diversity of the generated output can significantly drop. For future work, we would like to explore solutions including sampling theme representation with certain distribution and introducing new losses that penalize repetitive results.

\begin{figure}[t!]
 \centerline{
 \includegraphics[clip, trim=0cm 15cm 0cm 0.5cm, width = \columnwidth]{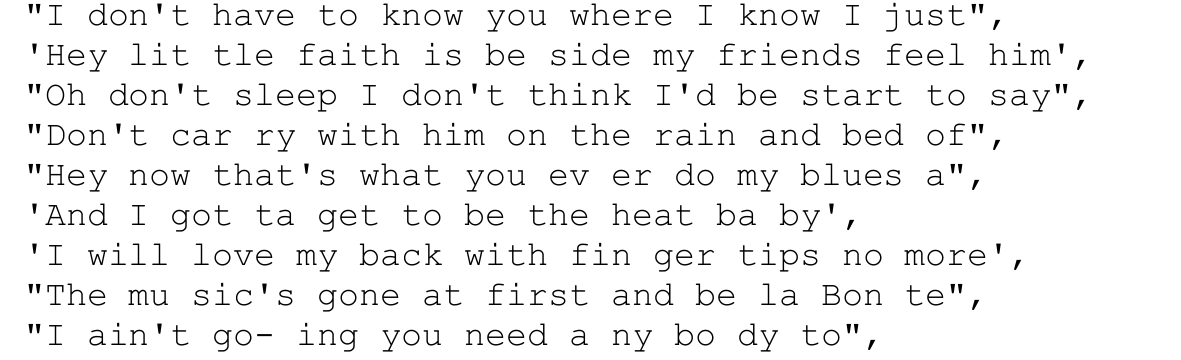}}
 \caption{Examples generated by SeqGAN.}
 \label{fig:SeqGAN_results}
\end{figure}

\begin{figure}[t!]
 \centerline{
 \includegraphics[clip, trim=0.5cm 17cm 0.5cm 3cm, width = \columnwidth]{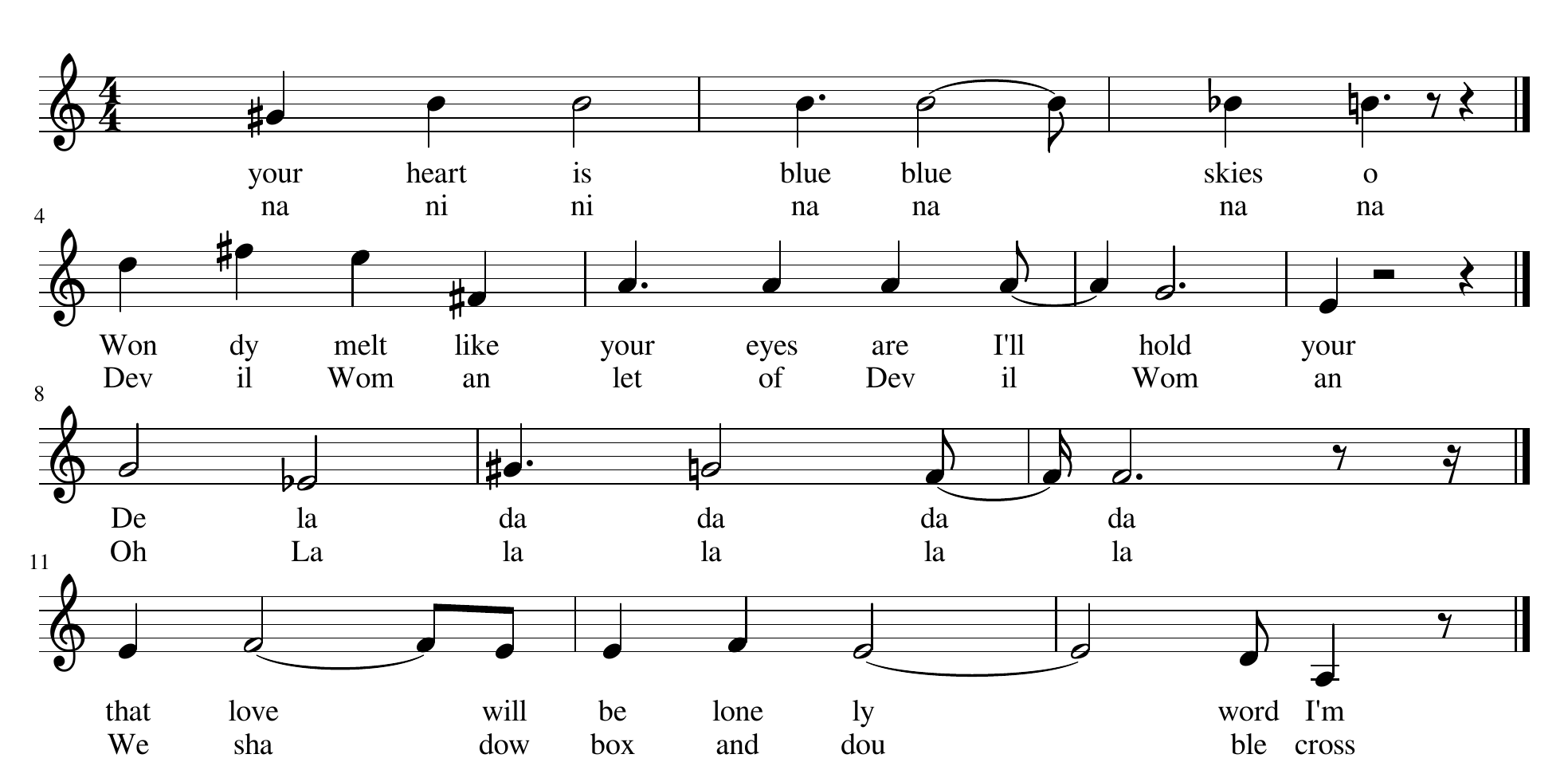}}
 \caption{Examples generated by MC-SeqGAN. Melody is shown on the staff. The first line in lyrics is the generated one, and the second line is the original lyrics.}
 \vspace{-10pt}
 \label{fig:MC_results}
\end{figure}

\subsubsection{Experiment 4}
\begin{figure}[t!]
 \centerline{
 \includegraphics[clip, trim=0.5cm 15.5cm 0.5cm 3cm, width = \columnwidth]{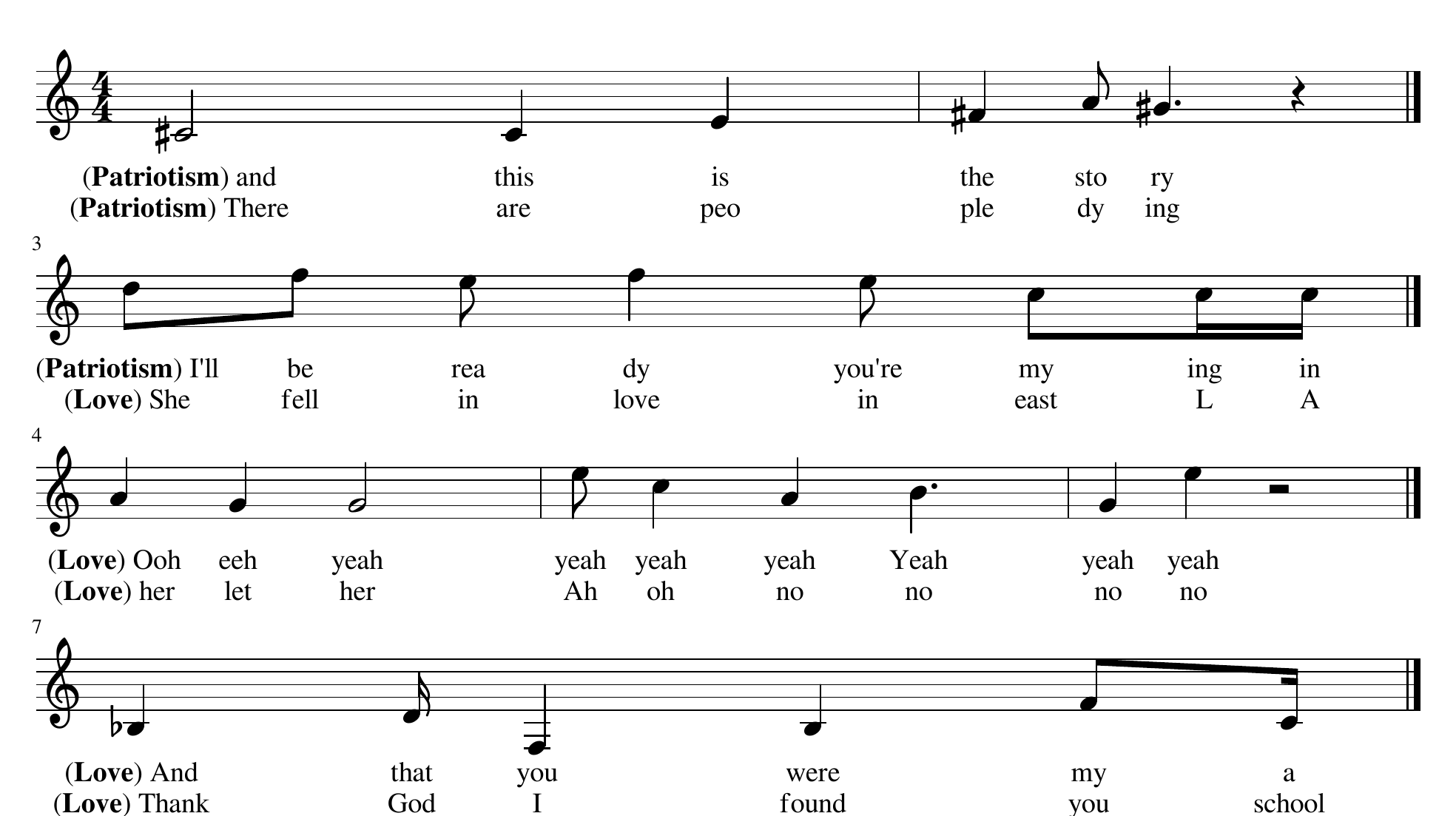}}
 \caption{Examples generated by TMC-SeqGAN. The preceding parenthesis indicate the extracted theme class}
 \label{fig:TMC_results}
\end{figure}

Figures~\ref{fig:SeqGAN_results}, \ref{fig:MC_results}, and \ref{fig:TMC_results} show example results for our SeqGAN based models. Most of the lyrics are human-readable, and their qualities are similar. Most generated syllables form an English word. For example, ``How long for-ev-er,'', and ``I'll be rea-dy.'' There are, however, some meaningless generated results such as ``Is it on su and some-one is there are'' and ``on your head I might come a.'' We also see some lyrics merely composed of ``placeholders'' such as ``Ah,'' and ``Ooh yeah.'' Even though this is common in lyrics, it might not be ideal as the output of a generative system. A possible post-processing step could filter out lines exclusively containing ``placeholders.'' 
An easily observable short-coming is that the generated lyrics do not have clear interrelations with each other. Some lines seem to be cropped from a longer sentence, such as ``a new child in the'', where the prepositional phrase is incomplete. This is directly related to our line-level data organization and random sampling and could be solved in the future by introducing longer multi-line sequences as inputs or include information from preceding line while generating the current line of lyrics. Given that SeqGAN suffers from a possible gradient vanishing problem, the improvement of interline relationships has to be investigated further.

The melody-conditioned lyrics are mostly singable and work generally well with the melody. Nevertheless, since the input melody and generated lyrics are short, it is hard to fully evaluate whether the content of the lyrics reflects the properties of the melody. For example, a melody in a minor scale might be better paired with sad lyrics. Moreover, it is common in composition and lyrics writing to match shorter notes with shorter syllables and longer ones with multi-syllabic words or vowels that can be intentionally prolonged \cite{pattison_1991_lyricswriting}. Even though this is not a hard rule and some violations have been witnessed in the training data, it could improve the flow of the lyrics and the perceived fit to the melody. Such restrictions are not enforced in the current system. Future work considering the syllable length could lead to interesting results.

\section{Conclusion}\label{sec:conclusion}

In this paper, we proposed a deep learning approach to lyrics generation conditioned on melodic information with SeqGAN architecture. We proposed an end-to-end system with melody and theme as conditioning input that utilizes a minimum amount of prior knowledge and pre-processing. We also have shown that the impact of the conditioning inputs on the quality of the generated lyrics is negligible and that their alignment with melodies is not impacted by the introduction of constraints, indicating the feasibility of adding more input conditions for enhanced user control of the generator. 
However, long-term text relation, rhyme, and syllabic structures learning have not yet been fully explored. These could be crucial for lyrics generation and should be studied in future work.  In addition, the utilization of the LDA algorithm output as pseudo ground truth for theme extraction is not ideal. A dedicated dataset with evenly distributed themes and human annotations would be helpful. Recently, large-scale pre-trained language models, e.g.\ GPT-3, have been shown to be potential in various text generation tasks \cite{brown2020language}. Incorporating melodic information to perform constrained lyrics generation with them will be worth exploring.






%
\bibliographystyle{IEEEtran}
\bibliography{MCGAN}

\end{document}